\documentclass[runningheads]{llncs}

\usepackage[utf8]{inputenc} 
\usepackage[T1]{fontenc}    
\usepackage{lmodern}

\usepackage{hyperref}       
\usepackage{url}

\usepackage{booktabs}       
\usepackage{amsfonts}       
\usepackage{nicefrac}       
\usepackage{microtype}      
\usepackage{xspace}

\usepackage{float}
\usepackage{tabularx}
\usepackage{algorithm}
\usepackage{algorithmic}

\usepackage{todonotes}
\usepackage[title]{appendix}

\usepackage{tikz}
\usetikzlibrary{positioning}
\usetikzlibrary{shapes.geometric,arrows.meta,decorations.markings}
\usetikzlibrary{calc,shapes.callouts}
\usepackage{wrapfig}
\usepackage{makecell}
\usepackage{array}

\usepackage{enumitem}
\setitemize{noitemsep,topsep=0pt,parsep=0pt,partopsep=0pt}

\usepackage[skip=2pt,font=scriptsize, textfont=it]{caption}

\usepackage{color, colortbl}
\definecolor{LightCyan}{rgb}{0.88,1,1}
\definecolor{Gray}{gray}{0.9}
\newcolumntype{a}{>{\columncolor{Gray}}c}

\def\systemname#1{\textsf{#1}\xspace}
\newcommand{\rlc}{\systemname{rlCoP}}
\newcommand{\plc}{\systemname{plCoP}}
\newcommand{\fc}{\systemname{fCoP}}
\newcommand{\lc}{\systemname{leanCoP}}
\newcommand{\flop}{\systemname{FLoP}}
\newcommand{\glc}{\systemname{graphCoP}}
\newcommand{\lazy}{\systemname{lazyCoP}}

\usepackage{subfig}

\usepackage{color}
\newboolean{showComments}
\setboolean{showComments}{true}
\ifthenelse
{\boolean{showComments}}
{
\newcommand{\wrk}[1]{\textcolor[rgb]{.0,.6,.0}{(\textsl{#1})}}
\newcommand{\evil}[1]{\textcolor[rgb]{.4,.0,.4}{(Evil reviewer: \textsl{#1})}}
}
{
\newcommand{\wrk}[1]{}
\newcommand{\evil}[1]{}
}

\usepackage{amsmath,amsfonts,bm}

\def\eqref#1{equation~\ref{#1}}

\def\1{\bm{1}}

\DeclareMathAlphabet{\mathsfit}{\encodingdefault}{\sfdefault}{m}{sl}
\SetMathAlphabet{\mathsfit}{bold}{\encodingdefault}{\sfdefault}{bx}{n}

\DeclareMathOperator*{\argmax}{arg\,max}

\begin{document}

\title{Towards Finding Longer Proofs}

\author{
  Zsolt Zombori\inst{1,2}
  \and Adri\'{a}n Csisz\'{a}rik\inst{1,2}
  \and Henryk Michalewski\inst{3,4}
  \and Cezary Kaliszyk\inst{5,3}
  \and Josef Urban\inst{6}
}

\authorrunning{Zs. Zombori et al.}

\institute{
  Alfr\'{e}d R\'{e}nyi Institute of Mathematics, Budapest
  \and E{\"o}tv{\"o}s Lor\'{a}nd University, Budapest
  \and University of Warsaw
  \and Google Inc.
  \and University of Innsbruck
  \and Czech Technical University in Prague
}

\maketitle

\begin{abstract}
We present a reinforcement learning (RL) based guidance system for
automated theorem proving geared towards Finding Longer Proofs
(\flop). Unlike most learning based approaches, we focus on
generalising from very little training data and achieving near complete
confidence. We use several simple, structured datasets with very long
proofs to show that \flop can successfully generalise a single
training proof to a large class of related problems. On these
benchmarks, \flop is competitive with strong theorem provers despite
using very limited search, due to its ability to solve problems that are
prohibitively long for other systems.

\end{abstract}

\section{Introduction}
Automated Theorem Proving (ATP) is the study of using machines for
formal mathematical reasoning. It is related to general game playing,
for example, the game of Go can be viewed as a simple formal system.
Building on the recent success of machine learning, a growing trend in
this field is to use learning methods to make theorem provers more
powerful. Several research projects have shown that learning can be
used to replace/surpass human-engineered heuristics. Despite huge
improvements, interesting mathematical theorems remain elusive
today. One crucial shortcoming of ATP systems is that they can typically find
only relatively short proofs.

In this paper, we address this shortcoming and ask the question of how
machine learning can be used to solve problems requiring very long
inference chains. We argue that the fundamental reason why current ATP
systems are limited to short proofs is that they focus on the search
aspect of the task in the space of inference steps. It is very natural
to see theorem proving as a search problem: each proof step involves a
choice from a set of valid inferences, yielding a search space that
grows exponentially with the length of the proof. Due to the
exponential blowup, the search is bound to fail beyond a certain depth
-- except for special classes of problems where one of the smart human
heuristics of the theorem prover allows for finding the solution
without a search. As W. W. Bledsoe observed \cite{bledsoe}:
``Automated theorem proving \dots is not the beautiful process we know
as mathematics. This is ‘cover your eyes with blinders and hunt
through a cornfield for a diamond-shaped grain of corn’.''

Approaches that try to avoid excessive search broadly fall into three
categories: 1) Perform large steps, such as the invocation of tactics,
decision procedures in SMT solvers~\cite{smt}, or other complex
algorithms. This approach is widely used in interactive theorem
provers, e.g. \cite{hollight,coq,isabelle}.  2) Perform hierarchical
reasoning by first creating a high-level proof plan and then gradually
refine it to the calculus level, e.g.
\cite{bundy_proofplanning,knowledge_based_proofplanning}. 3) Reason by
analogy, e.g. \cite{proof_by_analogy,analogy_Brock}.

Reasoning by analogy involves observing the proof of one problem,
extracting the core idea, and successfully applying it to
another. Note that using this formulation, success is barely dependent
on proof length. On the other hand, establishing mappings between
proofs is challenging and depends heavily on a proper data
representation, which has been from the beginnings of ATP a major
bottleneck for this approach. However, with the advent of machine
learning methods capable of automatically discovering good data
embeddings, the analogy approach seems worth revisiting.

In this work, we interpret analogical reasoning as building a model
that internalises a proof and then successfully applies it to a class
of related problems, without relying much on search. The trained
model is supposed to "know" the proof of an unseen, yet familiar
problem. This is a highly simplified approach which does not capture
the full potential of analogy, but we argue that it is a meaningful
start that will hopefully lead to more refined solutions. 
We select classes of problems where proofs are highly similar and
trained humans can often generalize a single demonstration to the
entire class, even if proof lengths greatly differ. We explore whether
machine learning can yield similar generalization.


Many successful ATP systems, such as 
\cite{malarea,JakubuvU19,enigma_ng,holist_exploration,rlcop,plcop,prop_invariant_embedding,openai_transformer_atp}
implement the MaLARea~\cite{malarea0,malarea} learning/reasoning loop
(described later also as the DAgger~\cite{dagger} meta-algorithm). The
MaLARea loop interleaves ATP runs based on the current models
(\emph{data collection phase}) with a \emph{training phase}, in which
these models are updated to fit the collected data.

An alternative family of reinforcement learning methods, including
Temporal Difference (TD) learning~\cite{Sutton1998}, continuously
update their models, allowing the system to bootstrap on itself. Such
methods have so far been mostly ignored by the theorem proving
community. In these methods, the search is usually replaced by
rollouts.  While MaLARea has been shown to yield good search
heuristics, we argue that rollout based data collection is more
suitable when the aim is to fully explore the space around a single
problem without overfitting to it. Our work has the following contributions.

\begin{itemize}
\item We introduce a new theorem proving algorithm \flop
  (Section~\ref{sec:algorithm}) based on a TD algorithm~\footnote{In
  particular, we use Proximal Policy Optimization~\cite{ppo} (PPO), a
  variant of the policy gradient method, which uses Temporal Difference 
  learning for optimization of the value function.} and the connection tableau
  calculus~\cite{connection_method}. \flop makes use of the
  curriculum learning algorithms presented by \cite{backplay} and
  \cite{single_demo_monte}. These techniques are well established in
  RL, however, they have never been applied to theorem proving before.
\item We introduce a synthetic dataset of increasingly difficult
  arithmetic problems, as well as two datasets from the Logical
  Calculi domain of the TPTP~\cite{tptp} library, augmented with
  lemmata (Section~\ref{sec:datasets}).
\item We show that when restricted to single shot evaluation --
  without search -- \flop performs very well, while another prover
  based on guided Monte Carlo Tree Search greatly degrades.
\item We evaluate \flop on our arithmetic benchmarks by training it on
  a single problem and show that it generalizes very well even when
  evaluated without search, allowing just a few proof attempts. This
  suggests that it has learned a simple form of reasoning by analogy.
\item We use the arithmetic benchmarks to compare \flop with
  state-of-the-art provers Vampire~\cite{vampire}, E~\cite{eprover},
  \lc~\cite{leancop} guided by human-designed strategies, and
  with \rlc~\cite{rlcop} -- an RL-based connection tableau prover. In
  the simple setup of unary encoding of numbers, \flop is only
  outperformed by a portfolio (multi-strategy) mode of a single
  manually optimized rewriting-based system and only after trying
  several of its autoconfiguration heuristics. When using binary
  encoding, \flop performs best, demonstrating its ability to
  generalize to long proofs.
\end{itemize}

Our datasets presented in Section~\ref{sec:datasets} seem to be
particularly suited for machine learning methods: some problems are
algorithmically simple, with long solutions and strong shared
structure (Robinson Arithmetic) while others are less similar, but
hierarchically structured (Logical Calculi). Nevertheless,
state-of-the-art systems struggle with solving some of the problems
(see Section~\ref{sec:experiments}). Furthermore, our problems are
much easier to analyze than typical heterogeneous proof corpora, hence
promising better understanding of the current limits of theorem
provers. The difficulty of our synthetic problems, as well as the
proof lengths, are easily adjustable, yielding a scalable RL benchmark
with interpretable failure modes (see Appendix~\ref{sec:failures}).

Our code, datasets and all experiment configuration files are
available at \url{http://bit.ly/code_atpcurr}\footnote{This
  distribution does not include the \fc theorem prover, which cannot
  yet be publicly released, however, a binary can be obtained upon
  request.}. Supplementary materials including screencasts with
gameplays performed in our environments are available at the project
webpage \url{http://bit.ly/site_atpcurr}.

\section{Related work}

\paragraph{Theorem Proving by Analogy.}
Analogy has long been considered one of the most important heuristics
in mathematical problem solving,
e.g.~\cite{polya,plausibleReasoning}. It also gained attention in
automated theorem proving, e.g.\cite{analogy_Brock,proof_by_analogy},
as an alternative of search-based methods. \cite{analogy_Brock} defines
analogical reasoning as ``the proof of one theorem is used to guide
the proof of a similar theorem by suggesting analogous steps''. They
rely on a user-provided matching between analogous concepts related to
the two theorems and try to reuse the proof steps (adjusted modulo
analogy) in the source proof during the construction of the target.
\cite{proof_by_analogy} aim to achieve this on a higher level of
abstraction by matching proof plans of a source and a target
problem. As the proof plan of the target problem is constructed, the
plan of the source is searched for steps that can be transformed into
a suitable step for the target. The set of allowable transformations
are predefined and designed for a narrow domain. For example, the
transformations given in ~\cite{proof_by_analogy} aim to carry a
result, such as the Heine Borel theorem, stated in ${\mathbb R}^1$
over to ${\mathbb R}^2$. The characteristic feature of these systems
is that search is performed on the meta level of plan mappings and
proof step transformations. The search space is often defined ad hoc
and is much smaller than that given by the inference rules of the
calculus.

A machine learning system that is trained to guide a theorem prover is
supposed to achieve a similar result, with two important improvements.
First transformations are learned, without the need for manual
engineering. Second, establishing mappings between proof steps (that
can be transformed into each other) should result from learning of
flexible and abstract features.  The flexibility and abstraction
allows for potentially reusing the same proof components several
times, as well as using components from different proofs, which goes
beyond earlier attempts that only establish direct matching between
the two proofs.

\paragraph{Machine learning systems for guiding theorem provers. }
A large body of research exists that aims to provide guidance for
theorem provers via machine learning. FEMaLeCoP~\cite{femalecop},
\rlc~\cite{rlcop,prop_invariant_embedding} \plc~\cite{plcop} and
\lazy~\cite{lazycop} guide the \lc~\cite{leancop} compact connection
tableau prover, which is also the system guided in our
project. Learning based guidance is added to the saturation based
E~prover~\cite{eprover} in
\cite{deep_guidance,enigma,enigma_ng,enigma_anonymous}. The HOList
project~\cite{holist,graph_holist,holist_exploration} builds guidance
on the tactic level\footnote{A tactic is a human-designed program
  which aggregates multiple proof steps.} for the HOL
Light~\cite{hollight} higher-order theorem prover. A distinctive
feature of all these systems is that they rely heavily on an external
search procedure, such as Monte Carlo Tree Search~\cite{uct}, or the
search engine of the guided prover. Learning is aimed at making search
more efficient and it is implemented in alternating iterations of
proof search and model fitting, according to the DAgger~\cite{dagger}
meta-algorithm, first used in MaLARea~\cite{malarea} for theorem
proving. In contrast with the above, we use an algorithm which uses
bootstrapping and
learns from generated rollouts, aiming to learn to generate entire
proof sequences in the \lc calculus. Such rollout based learning has
so far been barely used in theorem proving, with the noteable
exception of \cite{Crouse2019ADR}, developed in parallel with \flop,
which guides a saturation style prover using a simple policy gradient
RL algorithm.

Concurrently with our work,
\cite{DBLP:conf/mkm/PiotrowskiU20,DBLP:conf/mkm/UrbanJ20,openai_transformer_atp}
have used recurrent neural networks, attention and transformers to generate next proof
steps. E.g., \cite{openai_transformer_atp} report generalisation on
problems with relatively short proofs. In line with emphasizing
analogy over search, their evaluation protocol only allows for very
limited search in a single proof attempt~\footnote{A maximum of 4096
  search nodes are allowed.}.
Our work employs much smaller neural models and focuses on
generalizing to proofs with hundreds and thousands of steps (see
Figure~\ref{fig:flop_length}).

\paragraph{Provers guiding the \lc Connection Tableau Calculus. }
As noted above, a series of learning systems guide the \lc connection
calculus. Of these, we highlight three systems that use roughly the
same learning setup: \rlc~\cite{rlcop}, \plc~\cite{plcop} and
\glc~\cite{prop_invariant_embedding}. In these systems, the value and
policy functions of the guided MCTS algorithm are learned similarly to
\cite{thinking_fast_and_slow,alphazero}. \flop shares the same
manually developed features~\cite{ckjujv-ijcai15} with \rlc and \plc,
while \glc employs a graph neural network for feature extraction. We
use these systems as an important baseline in
Section~\ref{sec:experiments}. While the differences are important,
they play little role in our current investigation and we refer to
them jointly as \emph{mcts-CoP}s.

\section{\flop{} -- Main Algorithm}
\label{sec:algorithm}

\begin{figure}
    \centering
    \includegraphics[width=0.99\linewidth]{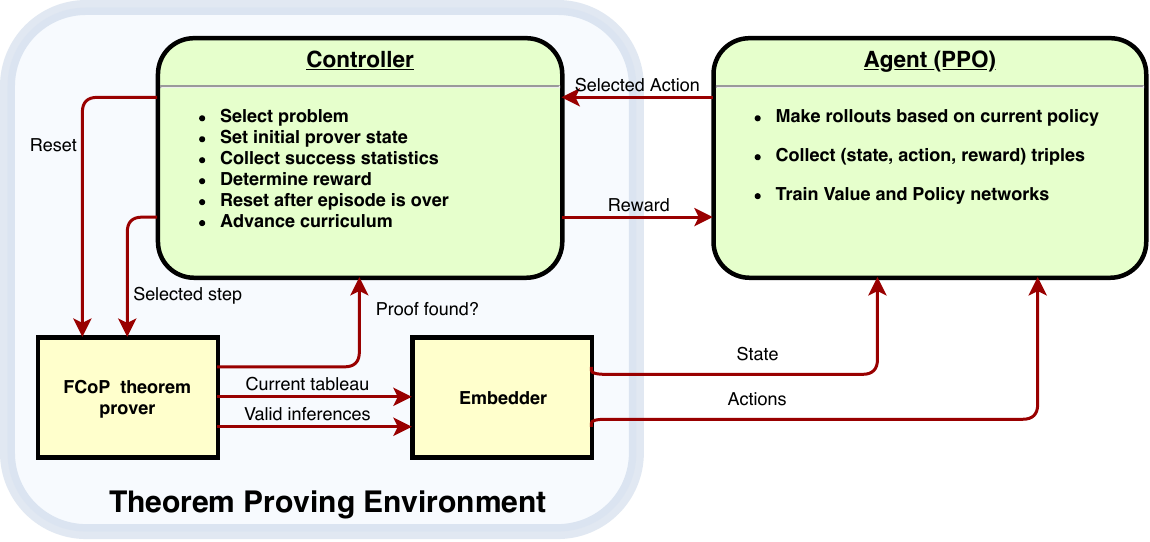}
    \caption{Theorem proving as a reinforcement learning environment.}
    \label{fig:architecture}
\end{figure}

\flop combines the connection tableau calculus with guidance based on
Temporal Difference and curriculum learning. A brief introduction to
the connection tableau calculus is provided in
Appendix~\ref{sec:foundations}. After each inference step, the prover
engine returns its current state as well as the set of valid actions,
i.e., valid inference steps that transform the current goal.  The
prover is encapsulated into a Reinforcement Learning (RL) environment. In
the following, we provide a brief summary of the relevant RL
techniques.

\subsection{Reinforcement Learning fundamentals}
The RL introduction below is highly selective, aiming to describe
Proximal Policy Optimization~\cite{ppo}, the method used in \flop. For
further details, we recommend \cite{Sutton1998}.

\paragraph{Markov Decision Process}
The mathematical foundation of the class of problems that
Reinforcement Learning aims to solve is given by Markov Decision
Processes (MDP). An $\mbox{MDP}(\mathcal{S}, \mathcal{A}, \mathcal{R},
\mathcal{P}, \gamma)$ describes a dynamic process and consists of the
following components: $\mathcal{S}$ is the set of states,
$\mathcal{A}$ is the set of actions, $\mathcal{R}:(\mathcal{S} \times
\mathcal{A}) \rightarrow \mathbb{R}$ is a reward function,
$\mathcal{P}: (\mathcal{S} \times \mathcal{A}) \rightarrow
\mathcal{S}$ is the state transition function and $\gamma$ is the
discount factor. We assume that an agent interacts with this MDP,
generating sequences of $(s_t, a_t, r_t)$ state-action-reward tuples,
called \emph{trajectories}. The agent is equipped with a \emph{policy}
function $\pi: \mathcal{S} \rightarrow \mathcal{A}$ which determines
which action it selects in a particular state. The aim of the agent is
to maximize its total accumulated reward
$\sum_{t \geq 0} \gamma^{t} r_t$. 
Several components of the model can be stochastic:
the reward function, the transition function, as well as the
policy. In such settings, the aim of the agent it to find the policy
$\pi^*$ that maximizes its cumulative expected reward, where future
rewards are discounted with the $\gamma$ discount factor:

$$ \pi^* = \argmax_{\pi} \mathbb{E} \lbrack \sum\limits_{t \geq 0}
\gamma^{t} r_t | \pi \rbrack$$

\paragraph{Policy Gradient}
One successful family of methods solves this task by considering a
parametric class of policy functions $\Pi = \{ \pi_\Theta, \Theta \in
\mathbb{R}^m \}$. We continuously sample trajectories from the current
policy and optimize the parameters $\Theta$ via gradient descent based
on the observed rewards. This is called \emph{policy gradient}, and
the RL literature contains numerous variants that differ in the
details of optimization.

One well known difficulty of policy gradient is the large variance in
the sampled trajectories, which makes convergence slow and requiring large
number of training samples. A popular technique to reduce variance is
to train a baseline model that esimates the expected reward from a
given state and optimize the policy with respect to the excess reward
on top of the baseline. This gives rise to the \emph{actor-critic
  framework}. We train two models jointly: a \emph{critic}: $V_\pi(s)$
that estimates the expected reward of trajectories starting from $s$
given policy $\pi$ and an \emph{actor}, which is our policy
$\pi$. Given some state $s$, we use the policy to sample an action
$a$. We then sample further transitions to estimate the expected
reward $Q_\pi(s, a)$ from state $s$ after taking action $a$. We define
\emph{advantage} as the difference between these two expectations:
$A_\pi (s, a) = Q_\pi(s, a) - V_\pi(s)$. Our optimization objective is then:

$$\min_{\Theta_V} \max_{\Theta_{\pi}} A_\pi (s, a)$$

where $\Theta_V$ and $\Theta_{\pi}$ are the parameters of the critic
and the actor, respectively. 

\paragraph{Proximal Policy Optimization} Policy gradient is an
\emph{on-policy} method, meaning that it optimizes the parameters of
the policy based on trajectories sampled from the same policy. In
contrast with \emph{off-policy} methods, which extract samples through
some other mechanism, policy gradient learning can be highly
unstable. This is because the change in policy potentially invalidates
the samples it was trained on. Proximal Policy Optimization~\cite{ppo}
(PPO) addresses this problem by introducing a soft constraint on the
magnitude of the policy updates. 

We maintain two instances of the
policy network: $\pi_{\Theta}$ which we aim to improve and
$\pi_{\Theta_{old}}$ which we sample from. The ratio
of the two policies gives us a measure of difference:

$$r_t(\Theta) = \frac{\pi_{\Theta}(a_t | s_t)}{\pi_{\Theta_{old}}(a_t
  | s_t)}$$

\noindent If this ratio lies outside of the range $[1-\epsilon,
  1+\epsilon]$, then the advantage function is clipped:

$$r_t^*(\Theta) = \mbox{clip}(r_t(\Theta), 1-\epsilon, 1+\epsilon)$$
$$A_{\pi_{\Theta}}^*(s, a) = \min(r_t(\Theta) A_{\pi_{\Theta_{old}}}, r_t^*(\Theta) A_{\pi_{\Theta_{old}}})$$

\noindent The two neworks are periodically synchronized to ensure that
they are not too different. PPO has been shown to strike a good
balance between simplicity and stability and is one of the most
popular policy gradient methods.

\subsection{Reinforcement Learning in \flop}
Theorem proving can be directly mapped into an MDP by treating prover
states as states, inference steps as actions and proof attempts as
trajectories. The only missing component is the reward function, which
we set to be

$$\mathcal{R}(s,a) =
\left\{
\begin{array}{ll}
  1  & \mbox{if perfoming a in s finishes the proof}\\
  0  & \mbox{otherwise}
  \end{array}
\right.
$$

\noindent Other reward functions are also possible, though we argue
that the selected one is most faithful to the task at hand: 1) we know
very little about progress before we have found a proof, hence the
zero reward for intermediary steps and 2) it is hard to tell if one
proof is better than another, hence the binary nature of the
rewards. 

\begin{figure}[htb]
  \begin{center}
    \includegraphics[width=0.9\linewidth]{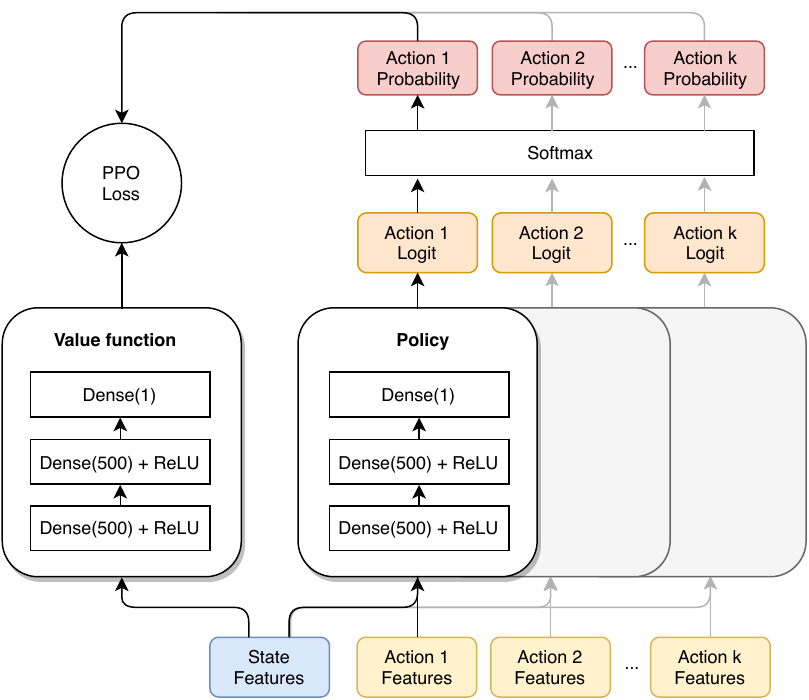}
  \end{center}
  \caption{Value and Policy network architectures in PPO. Their inputs
    are state and state-action pair features, respectively. The policy
    returns a score for each action, which are then normalized to a
    probability.}
  \label{fig:ppo_actions}
\end{figure}

Reward maximization directly corresponds to finding a proof. Hence, we
augment the core connection tableau calculus with a value (critic) and
a policy (actor) model trained using PPO. Classical proof search is
then replaced with generating proof attempts from the
policy. Figure~\ref{fig:architecture} shows the overall architecture
of the system and Figure~\ref{fig:ppo_actions} shows the policy and
value network architectures. The state and the actions (formulae) are
represented using previously developed features~\cite{ckjujv-ijcai15}.
The features include (suitably hashed) triples, pairs, and singletons
of adjacent nodes in the formula trees and the partial proof trees, as
well as some global features: number of open goals, number of symbols
in them, their maximum size and depth, length of the current path, and
two most frequent symbols in open goals. This means that the proof
states and the actions are presented as (sparse) fixed-length vectors.

\subsection{Curriculum Learning}

A fundamental challenge for an RL system that learns to prove theorems
from its own exploration is that rewards are sparse and binary. In
case proofs are long, this makes learning nearly impossible. To tackle
this, we use curriculum learning on the length of proofs in case a
proof is available. Initially, we start exploration from near the end
of the proof, making it easy to succeed and obtain positive reward. As
the system gets more confident, we gradually move the starting state
backwards along the given proof. This approach has already been
successfully applied in many RL experiments. When there is no good
alternative to the training proof, the system eventually learns those
steps, while random exploration helps to identify alternatives and
find novel proofs. Exploration also helps to learn steps that make the
proof impossible to finish. We can start learning with or without
training proofs. Each training problem can have its own curriculum
schedule, which can be restarted when a new proof is found. Curriculum
learning is an efficient tool for boosting rewards found during
exploration.

\subsection{Training algorithm}

Algorithm~\ref{algo:curriculum_on_proofs} gives an overview of the
learning loop.  First, in line 5 we sample a problem (in case there
are multiple). In lines 6--9 we interact with the prover and ensure
that its state corresponds to the one dictated by the current
curriculum. In lines 10--15 we generate a rollout (proof attempt)
iterating 1) prover steps, 2) featurization, and 3) sampling a next
action according to the policy. If a new problem is solved, we start
the curriculum on it in lines 17--19. If performance on a given
problem and curriculum reaches a threshold, we advance the curriculum
in lines 20--22. In line 24 we update the policy and value models
based on the specified number of episodes.

\begin{algorithm}[t]
\footnotesize
\begin{algorithmic}[1]
\REQUIRE problems $\mathcal{P}$, policy $\pi$, value $v$, train steps $\in \mathbb{N}$, threshold $\in [0..1]$, episodes between updates: k $\in \mathbb{N}$
\ENSURE trained policy $\pi$, trained value $v$, possibly proofs for some problems in $\mathcal{P}$ \\

\STATE $curriculum$ $\gets$  dictionary such that for each $p \in P$ with proof $Pr$ $curriculum[p] = \mbox{len}[Pr]-1$
\STATE steps $\gets 0$
\WHILE{steps < train steps}
    \FOR{j in 1..k}
    \STATE $p \gets$ random problem from problem set $\mathcal{P}$ \COMMENT{An episode corresponds to a problem}
    \STATE initialize prover on problem $p$
    \IF{$p$ has stored proof}
        \STATE Take $curriculum[p]$ proof steps according to stored proof
    \ENDIF
    \WHILE{ not episode over }
        \STATE $s', a'_1, a'_2 \dots a'_l \gets$ Query prover for current state and valid actions
        \STATE $s, a_1, a_2 \dots a_l \gets \mbox{feat}(s'), \mbox{feat}(a'_1), \mbox{feat}(a'_2) \dots \mbox{feat}(a'_l)$ \COMMENT{Extract features}
        \STATE Take action according to policy $\pi(a|s)$, observe reward $r$
        \STATE steps $\gets$ steps + 1
    \ENDWHILE    
    \STATE update success ratio for $p$
    \IF{$p$ is solved with proof $Pr$ and no proof of $p$ was known before}
        \STATE $curriculum[p] \gets \mbox{len}(Pr) -1$ \COMMENT{Start curriculum}
    \ENDIF
    \IF{success rate for $p$ > threshold}
        \STATE $curriculum[p] \gets curriculum[p] - 1$ \COMMENT{Advance curriculum}
    \ENDIF
    \ENDFOR
    \STATE Update policy $\pi$ and value $v$
 \ENDWHILE
\end{algorithmic}
\caption{\flop: Main Learning Loop}
\label{algo:curriculum_on_proofs}
\end{algorithm}

\subsection{Implementation details}
Most of the FLoP system is implemented in the Python programming
language, using the \cite{stable-baselines} RL framework. FLoP guides
the \fc~\cite{fcop} system, which is a reimplementation of \lc in the
OCaml programming language. The communication between the guidance and
prover components is provided via the C foreign language interface.

\section{Datasets}
\label{sec:datasets}

To evaluate our system, we select simple classes of theorems with
strong shared structure, giving a large room for learning-based
improvement. Our five datasets are described in
Table~\ref{tab:datasets}. The datasets are bundled into an
OpenAI-gym~\cite{openai_gym} compliant environment and can be tested
with modern RL algorithms.

\begin{table*}[htb]
  \caption{Three challenges defined in the theory of Robinson
    Arithmetic (RA)
  and two challenges from the Logical Calculi (LCL) domain of the TPTP
  library}
  \label{tab:datasets}
  \centering
    
  \bgroup
  \def\arraystretch{1.5}
  \small
  \begin{tabularx}{0.99\textwidth}{ p{0.13\textwidth}
      p{0.08\textwidth}  p{0.06\textwidth} p{0.6\textwidth} }
    \toprule 
    Name & Theory & Size & Description  \\
    \midrule
        {\bf RA-1} & RA & 1800 & Expressions of the form $N_1 +
        N_2 = N$, $N_1 \cdot N_2 = N$, where $0 \leq N_i < 30.$
        (Examples: $3\!+\!4\!=\!7$ or $5\!\cdot\!12\!=\!60$.) \\ 
        \midrule
        
        {\bf RA-2} & RA & 1000 & $T = N$, where $0 \leq N$, and
        $T$ is a random expression with $3$ operators and operands
        $N_i$ such that $0 \leq N_i < 10.$ (E.g.:
        $((3\!+\!4)\!\cdot\!2)\!+\!6\!=\!20$.) \\
         \midrule
        
        {\bf RA-3} & RA & 1000 & $T_1 = T_2$, where $T_1$ and $T_2$ are
        random expressions with $3$ operators and operands $N_i$ such
        that $2 \leq N_i < 10.$
        E.g. $((3\!+\!4)\!\cdot\!2)\!+\!6\!=\!((1\!+\!1)\!\cdot\!5)\!\cdot\!2$.)
        \\ 
        \midrule

        {\bf LCL-Eq} & LCL & 890 & TPTP domain: Logic Calculi
        (Equivalential) -- extended with lemmata from E prover. \\
        \midrule

        {\bf LCL-Imp} & LCL & 1204 & TPTP domain: Logic Calculi
        (Implication/Falsehood 2 valued sentential) -- extended with
        lemmata from E prover. \\ 
        \bottomrule

    \end{tabularx}
    \egroup
    \vspace{-5mm}
\end{table*}

Three datasets are built on the theory of Robinson
Arithmetic~\cite{robinson}, which defines addition and multiplication
on the nonnegative integers. Despite its relative simplicity, this
theory seems to be particularly suited for machine learning methods:
solutions are long and repetitive, while also challenging for
state-of-the-art systems (see Section~\ref{sec:experiments}). We examine
both unary (24 actions) and binary (40 actions) encoding of
numbers. The axioms of Robinson Arithmetic are given on the project webpage.
Increasing
the numbers in the conjecture greatly increases the length of the
proof, making this dataset suitable for detecting the length boundary
of various theorem provers.

Two datasets are extracted from the TPTP library, from the domain of
Logical Calculi with condensed detachment (LCL). These theorems have
been extensively studied from the early days of automated theorem
proving,
e.g. \cite{condensed_detachment,Peterson76,Kalman1978-JOHASS,Wos90b}. We
run E prover with a large time limit on the problems and augment the
dataset with lemmata extracted by E. As a result, many proofs of
simpler problems can be directly used as parts of the proofs of harder
problems. A direct analogy from one problem to the other is usually not
possible, however, shallow search is often sufficient to connect the
proofs of easier problems to the proof of harder ones. 

The LCL domain in the TPTP~\cite{tptp} library consists of statements
about various formal inference systems. LCL-Eq and LCL-Imp formalize
properties of the Equivalential Calculus and the Implication and
Falsum Calculus, respectively. Both are subsystems of the classical
propositional calculus, restricting the set of allowed connectives to
$\{\equiv\}$ and $\{\implies, \bot\}$. For both subsystems, the
appropriate variant of the \emph{condensed detachment} inference rule
($A, A \equiv B \vdash B$ and $A, A \implies B \vdash B$) constitutes
a \emph{strongly complete} inference system, i.e., whenever a formula
semantically follows from a set of premises, it also follows from the
set syntactically.
A number of complete axiomatizations of both the Equivalential
Calculus and the Implication and Falsum Calculus exist and the
theorems in our datasets establish connections between them.

All arithmetic problems in our dataset are quite simple for humans,
but in the case of logical calculi, some of the problems were posing a
challenge for mathematicians (see \cite{wos_1984_a_new_use}).


\section{Experiments}
\label{sec:experiments}

Our experiments with Robinson arithmetic aim to demonstrate that in
this highly structured dataset \flop is capable of extracting a
general proof pattern from one or two proofs and generalizing to
related proofs of arbitrary length, using a restricted few-shot
evaluation method (see below). Experiments~1, 2, and 3 compare \flop
with strong theorem provers using different fragments of the
arithmetic dataset, varying the complexity of the axiomatization
(unary vs. binary encoding of numbers) and the complexity of the
target theorems (RA-1, RA-2, RA-3). \flop is either the best or the
second-best in each experiment. In each of these experiments, \flop is
allowed $100$ proof attempts without backtracking: the first attempt
is a deterministic run with a high time limit ($1000$ sec) that
always selects the action maximizing the policy and the remaining $99$
runs are stochastic samples from the policy with a time limit of $60$
sec.

The LCL problems used in our experiments are less structured and
success is dependent on search, even if the hierarchical composition
of problems ensures that a relatively small search is sufficient to
generalize from easier problems to harder ones. Consequently, we
expect that search-based methods are better in this domain. However,
when search is completely disallowed during evaluation, we show in
Experiment~4 that \flop performs much better than the mcts-CoPs.
Finally, in Experiments 5 and 6 we demonstrate the benefit of using
curriculum learning.



\paragraph{Experiment 1: Comparison with other provers.}

\begin{wraptable}[10]{R}{0.5\textwidth}
  \vspace{-20pt}

    \caption{Comparing a random model, Vampire, E, leanCoP, \rlc and
      \flop, with respect to success ratio for RA-1, RA-2 and RA-3
      problems. Our method (\flop) is marked in grey. E$_1$ -- auto
      mode, E$_2$ -- auto-schedule mode, E$_3$ -- auto-schedule with
      renamed equality. The reason why \flop did not reach 100\% on
      RA-2 is that a few problems timeouted.}
    \label{tab:other_provers}
    \centering
    \resizebox{0.48\textwidth}{!}{
    \begin{tabular}{ l l l l l l l l a }
        \toprule
        Dataset & Random & Vampire & E$_1$ & E$_2$ & E$_3$ & leanCoP & \rlc & \flop \\
        RA-1 & $0.04$ & $0.60$ & $0.60$ & {\textbf{1.0}} & $0.54$ & $0.22$ & $0.86$ & {\textbf{1.0}} \\
        RA-2 & $0.05$ & $0.40$ & $0.39$ & {\textbf{1.0}} & $0.25$ & $0.14$ & $0.74$ & $0.99$ \\
        RA-3 & $0.00$ & $0.34$ & $0.28$ & {\textbf{1.0}} & $0.22$ & $0.01$ & $0.41$ & $0.67$ \\
        \bottomrule
    \end{tabular}}
\end{wraptable}

We compare \flop with a random model, two state-of-the-art
saturation-style theorem provers (E 2.4, Vampire 4.3.0), a heuristic guided
connection tableau prover (\lc 2.1), and \rlc (one of the mcts-CoPs).
Vampire, E, and \lc use human-designed strategies instead of
learning. We use these provers in the configuration used for CASC, the
yearly competition of fully automated theorem provers, employing a
time limit of $60$ sec. per problem. For E, we also report the results
of the \emph{auto-schedule} mode. For \rlc we used the hyperparameters
described in \cite{rlcop}, only modifying the policy temperature from
$2.5$ to $1.5$, as this works better with the Robinson datasets. The
number of inferences in MCTS was limited to $200000$. \rlc was trained
on the whole evaluation set, while \flop was trained on a single
problem: $1 \cdot 1 = 1$ and $1 \cdot 1 \cdot 1 = 1$ for RA-1 and
RA-2, respectively.\footnote{For a description of RA-3 training
  problems, see Experiment~2.} 

\begin{figure}[htb]
  \begin{center} 
    \includegraphics[width=0.45\textwidth]{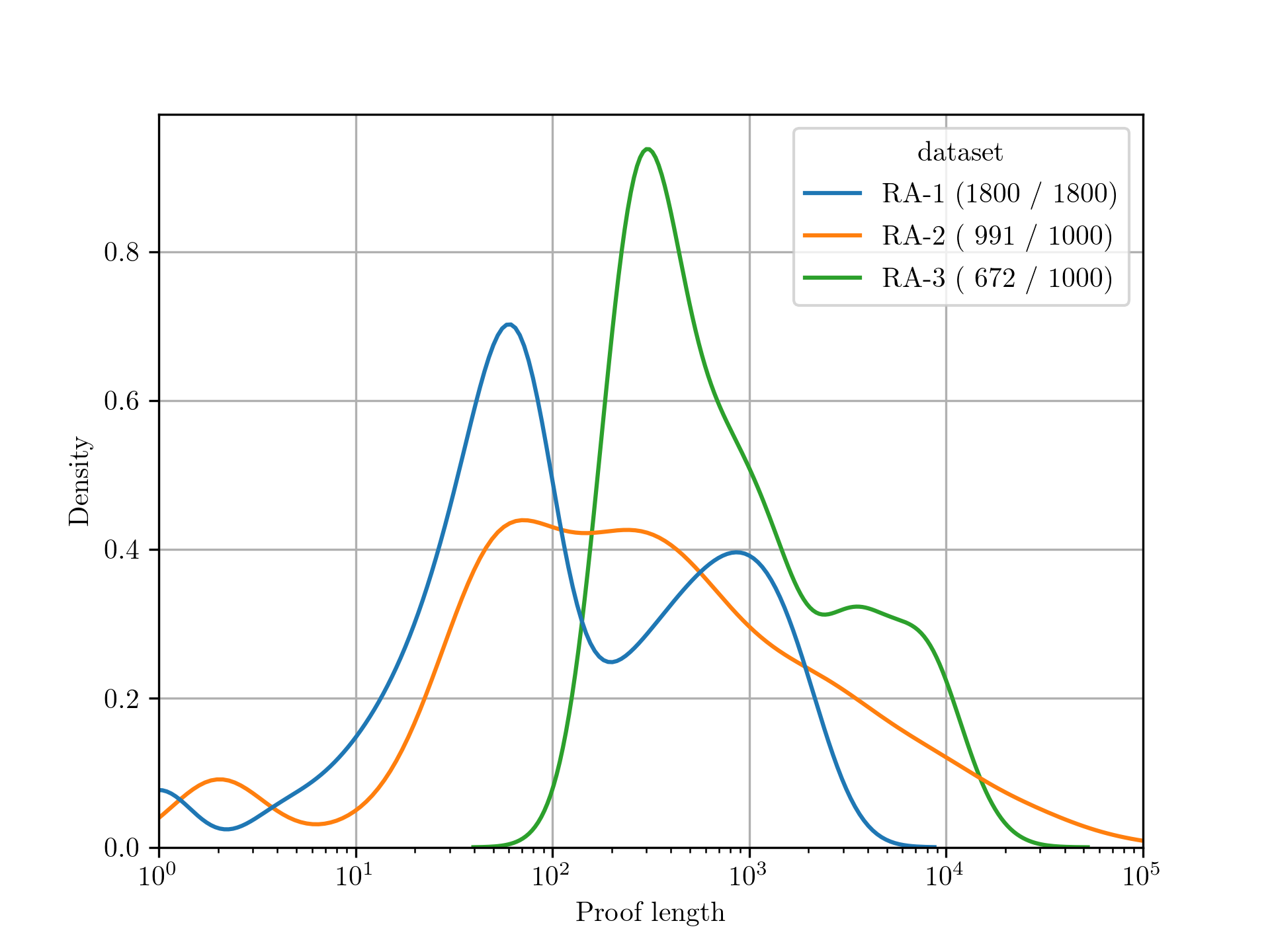}
    \includegraphics[width=0.45\textwidth]{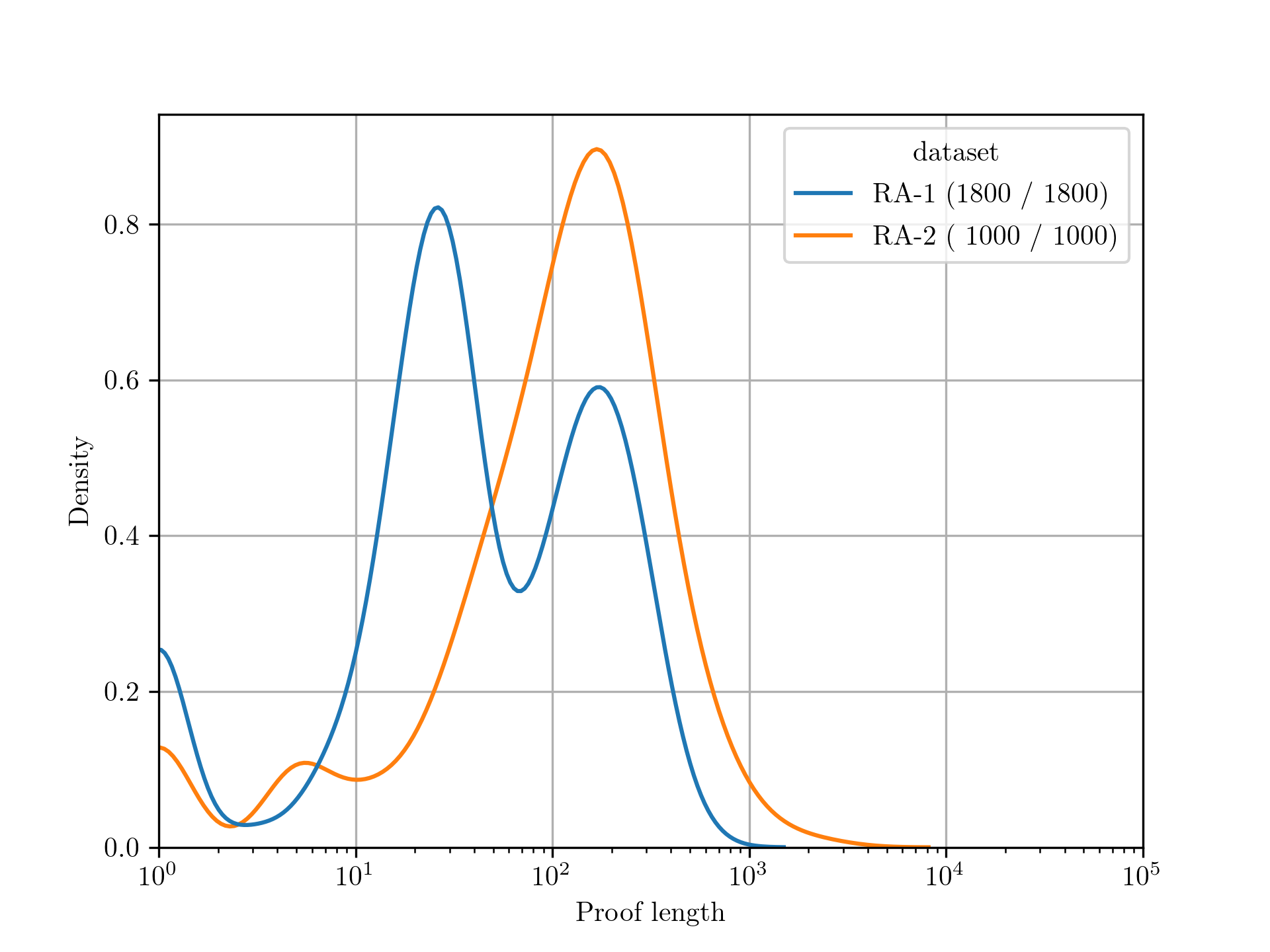}
  \end{center}
  \caption{Distributions of length of proofs found by \flop. Note the
    logarithmic scale. {\bf Left}: RA-1, RA-2 and RA-3 with average
    proof lengths 367, 2082, and 1864. {\bf Right}: binary RA-1 and
    binary RA-2 with average proof lengths 85 and 179.}
  \label{fig:flop_length}  
\end{figure}

Success ratios are given in Table~\ref{tab:other_provers}. \flop is
only outperformed by E's \emph{auto-schedule}, which tries multiple
strategies and finds one with the left-to-right ordering of all the
addition and multiplication axioms. This solves all of our problems
immediately without proof search by only rewriting to a normal
form~\cite{Rewriting}.  This demonstrates the power of equational
theorem proving when a suitable term ordering exists and can be found
by human-designed heuristics. This is, however, far from guaranteed in
general even in such simple domains, as witnessed by Vampire's failure
to find this ordering.  To evaluate E without access to its built-in
rewriting capability, we have renamed the equality to a new predicate
`eq' axiomatized exactly in the same way as in \lc. The auto-schedule
mode then becomes somewhat weaker than the auto mode, see
Table~\ref{tab:other_provers}.

\paragraph{Experiment 2: Harder Arithmetic Expressions.}

\begin{wraptable}[9]{R}{0.5\textwidth}
  \vspace{-20pt}
  \caption{Curriculum learning for RA-3 on two harder problems with
    proofs of $113$ and $108$ steps. We report success ratios and
    average proof lengths, based on $3$ runs. Standard deviations are
    given in parenthesis.}
  \label{tab:stage3_longer}
  \resizebox{0.48\textwidth}{!}{
    \begin{tabular}{ l l l }
      
      \toprule
      Training problem & Succ. & Len.\\
      \midrule
      $1 \cdot 2 + 1 + 1 = (1 + 1) \cdot 1 \cdot 2$ 
      & $0.32 (0.05)$ & $566 (14)$ \\
      \midrule
      $1 \cdot 2 + 1 + 1 = (1 + 1) \cdot 1 \cdot 2$ \\
      $(1 + 1 + 1) \cdot 2 = 2 \cdot 1 + 2 + 2$
      & {\textbf{0.67}} $(0.03)$ & $1864 (54)$ \\
      \bottomrule
  \end{tabular}}
\end{wraptable}

RA-3 consists of arithmetic equalities with random expressions on both
sides. This dataset is significantly more complex because there are
many ways of proving the same problem. Proofs are longer, too. For
\flop, we examined various training sets and found that the system is
very prone to overfitting. Most problems can be proven in many
different ways, that vary greatly in terms of how well they foster
generalization. It is true especially of easier problems that they can
be proven with ``shortcuts'' that hinder generalization (See more on
this in Appendix~\ref{sec:failures}).  The harder the problems, the
less likely they can be solved with such heuristic approaches, hence
harder training problems promise better training signal. We
demonstrate this by training \flop on a few harder problems with
proofs provided, making use of curriculum learning described in
Section~\ref{sec:algorithm}. A single longer training proof is
sufficient to yield meaningful generalization. Adding one more
training problem helps even more, as shows
Table~\ref{tab:stage3_longer}.

Figure~\ref{fig:flop_length} shows the distribution of the length of
proofs found by \flop. We can see that a large part of the problems
requires thousands of steps to solve, highlighting the need to avoid
search.

For \rlc, all RA-3 problems are too hard to solve without
guidance within the inference limit, so we started with the version
trained on the solutions of RA-2. Table~\ref{tab:other_provers}
shows that \flop is only outperformed by E's auto-schedule mode, which
again finds the rewrite ordering that solves all problems without search.

\paragraph{Experiment 3: Binary Number Encoding.}

\begin{wraptable}[7]{R}{0.5\textwidth}
  \vspace{-20pt}
    \caption{Comparing Vampire, E (auto-schedule mode), leanCoP, \rlc and \flop, using binary encoding of numbers.}
    \label{tab:binary_encoding}
    \centering
    \resizebox{0.48\textwidth}{!}{
    \begin{tabular}{ l l l l l a }
        \toprule
        Dataset & Vampire & E & leanCoP & \rlc & \flop \\
        RA-1 & $0.67$ & $0.81$ & $0.19$ & $0.56$ & {\textbf{1.0}} \\
        RA-2 & $0.62$ & $0.62$ & $0.13$ & $0.12$  & {\textbf{1.0}} \\
        \bottomrule
    \end{tabular}}
\end{wraptable}

We experiment with Robinson Arithmetic using binary encoding of
numbers. This makes the domain theory more complex: the total number
of actions increases from 24 to 40.~\footnote{Note that only a subset
  of these is applicable in a given state.} On the other hand,
proofs get shorter, as shows Figure~\ref{fig:flop_length}.
Again, we train \flop on a single proof: $3 \cdot 3=9$ and $(1 \cdot 2
+ 1) \cdot 3 = 9$ for RA-1 and RA-2, respectively.
Table~\ref{tab:binary_encoding} shows that provers get weaker, except
for Vampire and \flop. In particular, E is no longer capable of
solving the problems with rewriting only. \flop manages to generalize
from a single proof to the whole dataset despite the increased
action space and performs best in this experiment.

\paragraph{Experiment 4: Search vs. Eager Evaluation}

\begin{table}[htb]
  \caption{Comparing \flop and \plc using two different evaluation
    methods: 1) guided MCTS and 2) eager evaluation based on the
    policy model}
  \label{tab:flop_plcop_eval}
  \centering
  \begin{tabular}{l l l l l l}
    {\bf Prover} & {\bf Eval} & {\bf LCL-Eq} & {LCL-Imp} & {RA-1} & {RA-2} \\
    \hline
    \plc & MCTS & {\bf 47\%} & {\bf 61\%} & 65\% & 48\% \\
    \plc & Eager & 5\% & 5\% & 82\% & 49\% \\
    \flop & MCTS & 19\% & 24\% & 61\% & 31\% \\
    \flop & Eager & 19\% & 27\% & {\bf 100\%} & {\bf 99\%} \\
    \hline
  \end{tabular}
\end{table}


We compare \flop with \plc (one of the mcts-CoPs) using two different
evaluation methods. After training both systems on the whole dataset,
we evaluate them using 1) MCTS and 2) eager evaluation, i.e. always
select the action with the highest probability according to the policy
model. Table~\ref{tab:flop_plcop_eval} shows that \plc performs better
when search is allowed, especially for the more heterogeneous LCL
problems. However, \flop takes the upper hand in eager evaluation. For
the LCL problems, \plc collapses while \flop is unaffected. This
suggests that \plc depends heavily on the search procedure it used for
training. \flop cannot make good use of MCTS, which is somewhat
expected, since its policy and value networks were not trained for
that purpose. 
For the arithmetic datasets, both systems benefit from not doing
search because they reach proofs that are longer than what MCTS can
reach. For \flop, the removal of the depth limit reveals that it fully
mastered the two problem classes, regardless of depth.  The
performance of \plc gets even worse if the eager evaluation is based
on the value model, see Appendix~\ref{sec:eager_value}.  These results
are in line with our conjecture that the DAgger approach of \plc is
better for learning good search heuristics, while \flop is better at
internalizing a full proof pattern.

\paragraph{Experiment 5: Curriculum Learning vs only Exploration Based
  Learning.}

\begin{wraptable}[9]{R}{0.5\textwidth}
  \vspace{-20pt}
    \centering
    \caption{Curriculum Learning compared with only exploration based
      learning, on the LCL-Eq and LCL-Imp datasets, using 10M and 30M
      inference limit, respectively. We report the ratio of proofs
      found during training. The results are averages of 2 runs.}
    \begin{tabular}{ l l l }
        \toprule
        Dataset & Curriculum & No curriculum \\
        LCL-Eq & {\bf 0.24} (0) & 0.23 (0.001) \\
        LCL-Imp & {\bf 0.51} (0.002) & 0.45 (0.003) \\
        \bottomrule
    \end{tabular}
    \label{tab:lcl_curr_nocurr}
\end{wraptable}

When training proofs are not available, the positive reward signal only
occurs after the system solves a problem through
exploration. Afterward, curriculum learning ensures that the system
is continuously faced with a ``reasonably'' hard problem, alleviating
the sparse reward challenge of theorem proving.
We demonstrate this on the two LCL datasets. Here, before generating
each rollout, we randomly select a problem from the entire dataset. We
report the number of proofs found during training in
Table~\ref{tab:lcl_curr_nocurr}. Curriculum learning brings a small,
but consistent improvement when compared with only exploration-based
learning.

\paragraph{Experiment 6: Curriculum Learning vs. Supervised Learning}

\begin{wraptable}[10]{R}{0.5\textwidth}
  \vspace{-20pt}
  \centering
  \caption{Curriculum Learning vs Supervised Learning trained on
    proofs with extra steps added for distraction. \flop is barely
    affected, while supervised learning's performance
    degrades.}
    \resizebox{0.45\textwidth}{!}{
    \begin{tabular}{ l l l l }
    \toprule
    Data & Proof & Supervised & Curriculum \\
    & Lengths  & Succ. & Succ. \\
    \midrule
    RA-1 & $5$, $9$ & $0.98 (0.04)$ &\textbf{1(0.01)} \\
    & $9$, $11$ & $0.52 (0.08)$ & \textbf{0.98(0.01)} \\
    \midrule
    RA-2 & $5$, $9$, $23$ & \textbf{0.85(0.04)} & $0.76 (0.02)$ \\
    & $9$, $11$, $25$ & $0.59 (0.08)$ & \textbf{0.76(0.01)} \\
    \bottomrule
  \end{tabular}
    }
  \label{tab:supervised}
\end{wraptable}

When training proofs are available, a natural baseline of curriculum
learning is supervised learning on the proof steps. While such
behavioral cloning sometimes leads to great performance, we show in
Table~\ref{tab:supervised} that it greatly depends on the quality of
the given proof. We train RA-1 and RA-2 using the following sets of training problems:

\begin{enumerate}
  \item {\bf RA-1} $1+1=2$, $1\cdot 1=1$
  \item {\bf RA-2} $1+1=2$, $1\cdot 1=1$, $1 \cdot 1 \cdot 1 = 1$
\end{enumerate}

We take the ``nice'' proofs ($5$, $9$ and $23$ steps) of these
problems and construct variants with $2$-$3$ extra steps added. We
observe that supervised learning degrades as superfluous steps are
introduced, while \flop's exploration allows the system to recover and
find the original proofs.

\section{Conclusion and Future Work}

We have built \flop, a proof guidance system based on a variant of
temporal difference reinforcement learning, addressing the problem of
finding long proofs in an exponential search space. Previous work
\cite{hints,loops} focused on finding long proofs with the help of
human-designed heuristics. We showed that \flop is capable of
extracting proof patterns via learning and can generalise to much longer
proofs, implementing a simple form of reasoning by analogy.
We believe that mastering analogical reasoning is an important step in creating human-level automated mathematicians.
We presented a set of theorem proving
datasets that are suitably challenging for existing learning methods
and are intended to become a general-purpose testing ground for
reinforcement learning methods. We showed that
\flop can outperform strong theorem provers on some of these datasets.
We find that curriculum learning is a useful component of the learning
algorithm as it allows for amplifying training signal when proofs are
long.

In this paper, we focused on extracting a single proof pattern during
training. A natural continuation will be to extract a portfolio of
patterns from a larger pool of training problems. Transformer
models are promising tools to achieve this, given their recent success
in mastering several tasks in parallel. Transformers might
also be capable of producing large chunks of proofs in a single
inference step.

\section{Acknowledgments}
Adri\'{a}n Csisz\'{a}rik and Zsolt Zombori were supported by the
European Union, co-financed by the European Social Fund
(EFOP-3.6.3-VEKOP-16-2017-00002), the Hungarian National Excellence
Grant 2018-1.2.1-NKP-00008 and by the Hungarian Ministry of
Innovation and Technology NRDI Office within the framework of the
Artificial Intelligence National Laboratory Program. The work of
Henryk Michalewski was supported by the Polish National Science Center
grant UMO-2018/29/B/ST6/02959.  Cezary Kaliszyk was supported by ERC
grant no.\ 714034 \textit{SMART}.  Josef Urban was supported by the
\textit{AI4REASON} ERC Consolidator grant number 649043, and by the
Czech project AI\&Reasoning CZ.02.1.01/0.0/0.0/15\_003/0000466 and the
European Regional Development Fund.

This research was supported by the PL-Grid Infrastructure. In particular, quantitative results of \flop reported in this paper were performed using the Prometheus supercomputer, located in the Academic Computer Center Cyfronet in the AGH University of Science and Technology in Kraków, Poland.

\bibliography{atpcurr}
\bibliographystyle{splncs04}

\newpage
\appendix




\section{Robinson Arithmetic}
\label{sec:robinson}

Robinson Arithmetic defines basic properties of arithmetic expressions
on the nonnegative integers. The signature of the language contains
atom '$o$' (representing $0$), functions '$s$', '$\mbox{plus}$' and
'$\mbox{mul}$' (for $+1$, $+$ and $\cdot$, respectively), and the
equality predicate '$=$'. For example, formula $2 \cdot 1 + 1 = 3 + 0$
is written as
$$plus(mul(s(s(o)), s(o)), s(o)) = plus(s(s(s(o))), o).$$

\begin{table}[htb]
    \caption{Axioms of Robinson Arithmetic extended with special axioms for handling equality.}
    \label{tab:axioms}
    \centering
    \resizebox{0.98\textwidth}{!}{
    \begin{tabular}{ l l }
        \toprule
        Name & Axiom \\
        \midrule
        zero successor & $\forall X: \neg (o = s(X))$ \\
        different successors & $\forall X, Y: (s(X) = s(Y)) \Rightarrow (X = Y)$ \\
        plus zero & $\forall X: plus(X, o) = X$ \\
        plus successor & $\forall X, Y: plus(X, s(Y)) = s(plus(X, Y))$ \\
        mul zero & $\forall X: mul(X, o) = o$ \\
        mul successor & $\forall X, Y: mul(X, s(Y)) = plus(mul(X, Y), X)$ \\
        \midrule
        equality reflexivity & $\forall X: X=X$ \\
        equality symmetry & $\forall X, Y: (X = Y) \Rightarrow (Y = X)$ \\
        equality transitivity & $\forall X, Y, Z: (X = Y) \land (Y=Z) \Rightarrow (X = Z)$ \\
        congruence of s & $\forall X, Y: (X = Y) \Rightarrow (s(X) = s(Y))$ \\
        congruence of plus & $\forall X_1, X_2, Y_1, Y_2: (X_1 = X_2) \land (Y_1 = Y_2) \Rightarrow plus(X_1, Y_1) = plus(X_2, Y_2)$ \\
        congruence of mul & $\forall X_1, X_2, Y_1, Y_2: (X_1 = X_2) \land (Y_1 = Y_2) \Rightarrow mul(X_1, Y_1) = mul(X_2, Y_2)$ \\
        \bottomrule
    \end{tabular}}
\end{table}

We use the axioms provided in Table~\ref{tab:axioms}. The table also
contains special axioms that are added by \lc to handle the equality
predicate.  The unary representation of numbers (e.g., $s(s(s(o)))$
represents $3$) results in large expressions and long proofs as the
numbers increase. For example, $((8+5) \cdot 8) \cdot 5 = 520$ takes
over $16000$ steps to prove in \fc. We show an example of such proof
on the project website. The total number of actions if 24.

\section{Robinson Arithmetic with Binary Encoding}
\label{sec:robinson_binary}

\begin{table}[htb]
    \caption{Axioms of Robinson Arithmetic using binary encoding of numbers, extended with special axioms for handling equality.}
    \label{tab:axioms_binary}
    \centering
    \resizebox{0.98\textwidth}{!}{
  \begin{tabular}{ p{0.2\textwidth} p{0.9\textwidth} }
        \toprule
        Name & Axiom \\
        \midrule
        zero successor & $\forall X, Y: \neg (n0 = b(X,Y))$ \\
        one successor & $\forall X, Y: \neg (n1 = b(X,Y))$ \\
        different successors & $\forall X_1, X_2, Y_1, Y2: (b(X_1, Y_1) = b(X_2, Y_2)) \Rightarrow ((X_1 = X_2) \land (Y_1 = Y_2))$ \\
        predecessor & $\forall X: (X = n0) \lor (X = n1) \lor (\exists Y, Z: b(Y,Z) = X)$ \\
        plus zero & $\forall X: plus(X, n0) = n1$ \\
        plus one1 & $plus(n0, n1) = n1$ \\
        plus one2 & $plus(n1, n1) = b(n0,n1)$ \\
        plus one3 & $\forall X: (plus(b(n0,X),n1) = b(n1,X))$ \\
        plus one3 & $\forall X: (plus(b(n1,X),n1) = b(n0,plus(X,n1)))$ \\
        plus more1 & $\forall X, Y: (plus(n0,b(X,Y)) = b(X,Y))$ \\
        plus more2 & $\forall X, Y: (plus(n1,b(X,Y)) = plus(b(X,Y), n1))$ \\
        plus more3 & $\forall X_1, Y_1, X_2, Y_2: (plus(b(X_1,Y_1),b(X_2,Y_2)) = plus(b(X_1,plus(Y_1,Y_2)),X_2))$ \\
        mul zero1 & $\forall X: (mul(X,n0) = n0)$ \\
        mul zero2 & $\forall X: (mul(n0,X) = n0)$ \\
        mul one1 & $\forall X: (mul(X,n1) = X)$ \\
        mul one2 & $\forall X: (mul(n1,X) = X)$ \\
        mul more & $\forall X_1, Y_1, X_2, Y_2: (mul(b(X_1, Y_1),
        b(X_2, Y_2)) = $ $plus(plus(plus(b(n0,b(n0,mul(Y_1, Y_2))),  b(n0, mul(Y_1, X_2))),$ $b(n0, mul(X_1, Y_2))), mul(X_1, X_2)))$ \\
        \midrule
        equality reflexivity & $\forall X: X=X$ \\
        equality symmetry & $\forall X, Y: (X = Y) \Rightarrow (Y = X)$ \\
        equality transitivity & $\forall X, Y, Z: (X = Y) \land (Y=Z) \Rightarrow (X = Z)$ \\
        congruence of b & $\forall X_1, X_2, Y_1, Y_2: (X_1 = X_2)
        \land (Y_1 = Y_2) \Rightarrow (b(X_1,Y_2) = b(X_2,Y_2))$ \\
        congruence of plus & $\forall X_1, X_2, Y_1, Y_2: (X_1 = X_2) \land (Y_1 = Y_2) \Rightarrow plus(X_1, Y_1) = plus(X_2, Y_2)$ \\
        congruence of mul & $\forall X_1, X_2, Y_1, Y_2: (X_1 = X_2) \land (Y_1 = Y_2) \Rightarrow mul(X_1, Y_1) = mul(X_2, Y_2)$ \\
        \bottomrule
    \end{tabular}}
\end{table}

Using binary representation makes the theory of Robinson Arithmetic
more complex. Constant symbols $n0$ and $n1$ stand for $0$ and $1$,
while term $b(X,Y)$ represents $X + 2 \cdot Y$. For example, $10$ is
represented as $b(n0,b(n1,b(n0,n1)))$. We can see in
Table~\ref{tab:axioms_binary} that the number of axioms increases from
$6+6$ to $17+6$. Correspondingly, the number of actions during proof
search increases from 24 to 40.

\section{Experiment hyperparameters}
\label{sec:hyperparams}

Our hyperparameters were selected using small grid searches. We
checked standard RL parameters (e.g., the discount factor), parameters
related to curriculum scheduling (e.g., local vs. global), neural
network architectures (1--5 layers with 128--1024 neurons), feature
sizes (64--1024) and training steps ($10^5$ -- $10^8$). Parameters
used in the experiments are described in configuration files which are
accessible along with the shared codebase.

\section{Eager evaluation based on the value model}
\label{sec:eager_value}

\begin{wraptable}{R}{0.5\textwidth}
  \vspace{-20pt}
  \caption{Comparing \plc using MCTS and two different eager evaluation
    methods based on hte policy and value models}
  \label{tab:plcop_eval}
  \centering
  \resizebox{0.48\textwidth}{!}{
    \begin{tabular}{l l l l l}
      {\bf Dataset} & {\bf Prover} & {\bf MCTS} & {\bf Eager Policy} & {\bf Eager Value} \\
      \hline
      LCL-Eq & \plc & {\bf 47\%} & 5\% & 1\%  \\
      \hline
      LCL-Imp & \plc & {\bf 61\%} & 5\%  & 1\% \\
      \hline
      RA-1 & \plc &  65\% & {\bf 82\%} & 3\% \\
      \hline
      RA-2 & \plc  & 48\% & {\bf 49\%} & 5\% \\
      \hline
    \end{tabular}}
\end{wraptable}

Given that the evaluated RL algorithms train both a policy and a value
model, an alternative of policy-based eager evaluation is to select
the action whose successor state has the highest value
score. Table~\ref{tab:plcop_eval} shows, however, that the value-based
evaluation is much worse for each dataset. We conjecture that this is
because assigning a value to a never observed state is much harder
than selecting from a smaller set of actions.

\section{Failure Modes}
\label{sec:failures}

Despite the apparent simplicity of our arithmetic learning
environments, a learning system aiming to solve them has to overcome
some hard challenges.  We have decided to describe these challenges in
detail as they are present in other domains as well, even if it may be
harder to detect.

{\bf Failure type 1.}  The reward mechanism of our RL system is biased
towards shorter proofs. However, many problems have ``shortcuts'' that
allow for shorter proofs, but that do not generalize well. Consider
formula $(1+1) + (2\cdot 2) = (0+2) + 4$. There are two ways to prove
this equality: 1) compute the values of the expressions on both sides
of the equation and notice that they are the same or 2) show that $1+1
= 0+2$ and $2\cdot 2 = 4$. The former generalizes better, but the
latter results in a shorter proof. Hence, training on this problem
might negatively affect the performance of the prover. This is what
prevents \flop to bootstrap itself in RA-3, i.e., train on easy
problems and generalize to harder ones. We find that providing some of
the harder problems (having longer proofs) helps to avoid misleading
shortcuts.


{\bf Failure mode 2.}  \fc features do not take into account the order
of the arguments of a function, i.e., $f(a,b)$ and $f(b,a)$ have the
same features. This is problematic for RA-3, since $A = B$ and $B
= A$ require different inferences. We addressed this problem by 1)
extending state features with those of the preceding action as a
substitute of memory, 2) modified the features to include argument
order.

{\bf Failure mode 3.}  Some ''rare'' events are hard to generalize
because the system sees very few relevant samples during
training. This is the case with applying commutativity of equality
(replacing $A=B$ with $B=A$), which is only required in RA-3 and
ideally only once per proof when we move the focus from one side of
the equation to the other. In Experiment~4, when we trained on a
single longer proof, we have noticed that the system was very unsure
about this action which resulted in many failed proof attempts. Adding
another training proof was enough to overcome this and the success
score increased from $32\%$ to $67\%$.

\section{The \lc Connection Tableau Calculus}
\label{sec:foundations}

\begin{figure}[htb]
  \begin{center}
    \includegraphics[width=1.0\linewidth]{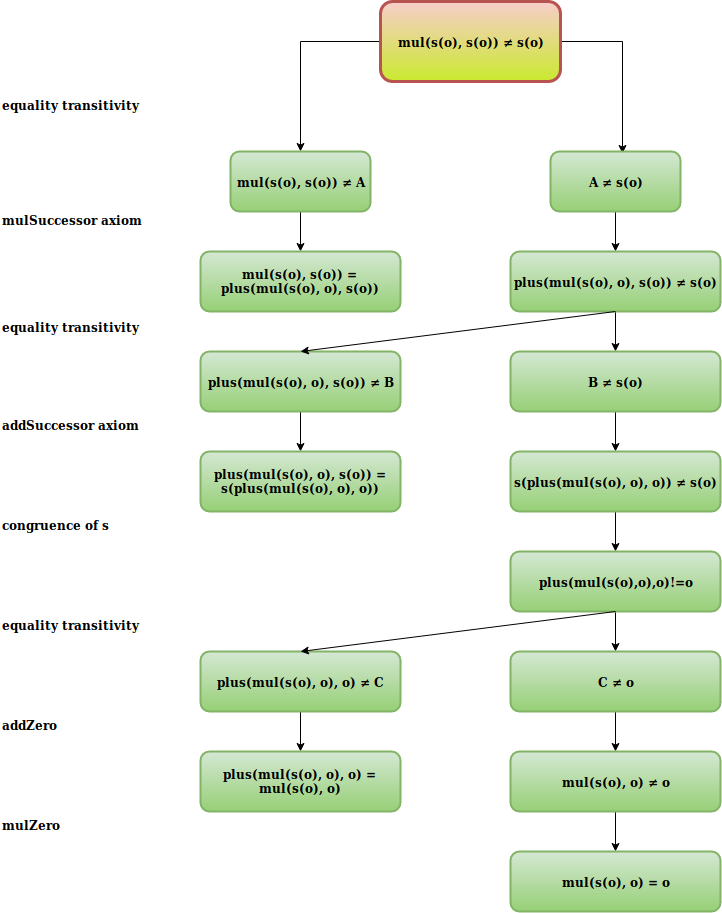}
  \end{center}
  \caption{A closed tableau tree of the proof of $1 \cdot 1 = 1$. On
    the left we list the actions taken in the proof. See {\normalfont
      \url{http://bit.ly/site_atpcurr}} for details.  }
  \label{fig:tableau}
\end{figure}

\flop provides guidance for of the very compact \lc~\cite{leancop}
connection tableau calculus. The calculus was originally implemented
in Prolog, but it also has an OCaml reimplementation \fc~\cite{fcop}
and \flop can be used to guide both systems.

We briefly describe the connection tableau calculus, assuming basic
first-order logic and theorem proving
terminology~\cite{HandbookOfAR}. The input is a (mathematical)
problem consisting of \emph{axioms} and \emph{conjectures} formally
stated in first-order logic (FOL). The calculus searches for
\emph{refutational proofs}, i.e. proofs showing that the axioms
together with the negated conjectures are \emph{unsatisfiable}. The
FOL formulas are first translated to \emph{clause normal form} (CNF),
producing a set of first-order \emph{clauses} consisting of
\emph{literals}, e.g. $\{\forall X, Y: (f(X) | r(X,Y | \lnot f(Y)),
f(a)\}$. Proof search starts with a \emph{start clause} as a
\emph{goal} and proceeds by building a connection tableau by
repeatedly applying \emph{extension steps} and \emph{reduction steps}.

The extension step connects (\emph{unifies}) the \emph{current goal}
with a complementary literal of a new clause. This extends the
\emph{current branch}, possibly splitting it into several branches if
there are more literals in the new clause, and possibly
\emph{instantiating} some variables in the tableau.  The reduction
step connects the current goal to a complementary literal of the
\emph{active path}, thus \emph{closing} the current branch. The proof
is finished when all branches are closed. The extension and reduction
steps are nondeterministic, requiring backtracking in the standard
connection calculus. Brute force search such as \emph{iterative
  deepening} can be used to ensure completeness.
Figure~\ref{fig:tableau} shows a \emph{closed connection tableau},
i.e., a finished proof tree where every branch contains
\emph{complementary literals} (literals with opposite polarity). This
shows that the set of clauses is unsatisfiable.

\lc represents theorem proving as a one-person game. The game ends
with success if a proof is found. The prover has many choices to make
along the way, in particular it can select from several valid
extension and reduction steps. Whether a step is valid depends on the
unification condition, i.e., if the current goal unifies with the
negation of a literal in the corresponding clause. The full
information about the game state consists of all previous proof steps,
the partial proof tree (proof state) and the current goal.

The search space of the prover is exponentially large in the length of
the proof. In \lc, the action space is roughly correlated with the
size of the axiom set. While this can be large for large problems,
typically only a few actions are available in any particular state.

\end{document}